\documentclass[twocolumn,prl,showpacs]{revtex4}
\usepackage{graphicx}
\usepackage{dcolumn}
\usepackage{bm}
\usepackage{amsmath}

\begin{document}

\preprint{}
\title{Giant spin rotation in the normal metal/quantum spin Hall junction}
\author{Takehito Yokoyama$^{1}$, Yukio Tanaka$^{1}$, and Naoto Nagaosa$^{2,3}$}
\affiliation{$^1$Department of Applied Physics, 
Nagoya University, Nagoya, 464-8603,
Japan \\ 
$^2$ Department of Applied Physics, University of Tokyo, Tokyo 113-8656, Japan \\
$^3$ Cross Correlated Materials Research Group (CMRG), ASI, RIKEN, WAKO 351-0198, Japan 
}
\date{\today}

\begin{abstract}
We study theoretically reflection problem in the junction between a normal metal and an insulator 
characterized by a parameter $M$, which is a usual insulator for $M>0$ or a 
quantum spin Hall system for $M<0$. The spin rotation angle $\alpha$ at the reflection is obtained in the plane of $M$ and the incident angle $\theta$ measured from the normal to the interface. The $\alpha$ shows rich structures around the quantum  critical point $M=0$ and $\theta=0$, i.e., $\alpha$ can be as large as $\sim \pi$ at an incident angle in the quatum spin Hall case $M<0$ because the helical edge modes resonantly enhance the spin rotation, which can be used to map the energy dispersion of the helical edge modes.  
As an experimentally relevant system, we also study spin rotation effect in quantum spin Hall/normal metal/quantum spin Hall trilayer junction.
% It is clarified that multiple reflection in this junction leads to the spin rotation even when incident angle averaged.
%Even more enhanced spin rotation can be obtained by the multiple reflection.

\end{abstract}

\maketitle

%--- title ---

%--- author ---

%
%--- address ---

%
%--- date ---

% It is always \today, today,
%  but any date may be explicitly specified
%-----------------------------------------------------------
%   Abstract
%-----------------------------------------------------------

%-----------------------------------------------------------

% PACS, the Physics and Astronomy
% Classification Scheme.
%\keywords{Suggested keywords}%Use showkeys class option if keyword
%display desired
%\section{Introduction}
%-----------------------------------------------------------

The ultimate goal of the spintronics is to manipulate the spins without the magnetic field, and the relativistic spin-orbit interaction is the key to realize it. One example is the Datta-Das spin transistor \cite{Datta} where the spin-orbit interaction modified by the gate voltage controls the rotation of the spins of the carriers. However, usually the strength of the spin-orbit interaction is weak and its influence is small especially in semiconductors. Nevertheless, its effects are now clearly observed experimentally, e.g., in the Rashba spin splitting \cite{Nitta}, and spin Hall effect (SHE) \cite{DP,Murakami,Kato,Wunderlich}. 
Further enhancement of the spin-orbit effects in semiconductors is highly desired, and for that purpose we propose in this Letter to use the quantum spin Hall (QSH) system\cite{Mele,Bernevig} and its junction to the normal metal (N) as the spin rotator with a giant angle $\alpha$ of the order of $\pi$ due to the resonance with the helical edge channels. This effect can be used to map the energy dispersion of the helical edge modes, and also to produce the spin current parallel to the interface with the average over the incident angle.

%The SHE was originally proposed long ago as caused by the skew scattering analogous to the anomalous Hall effect in ferromagnetic metals \cite{DP}. Recent developments are triggered by the theoretical proposal for the intrinsic mechanism of SHE, where the Berry curvature in the momentum space gives the spin Hall conductivity.\cite{Murakami} This idea has lead to the discovery of the QSH system and the topological insulator, which is characterized by a new topological number ($Z_2$ number) corresponding to the number of the helical edge channels at the edge of the sample.\cite{Mele,Bernevig} 
Naively, QSH system can be regarded as the two copies of the integer quantum Hall systems for up and down spins with the opposite chiralities. Hence, the chiral edge modes for up and down spins with the opposite propagating directions are expected, which is called the helical edge modes.\cite{wu2006,xu2006,Fu,Qi} Two sets of the helical edge modes can be mixed by e.g., the impurity scattering, and open the gap, but when the number of these sets is odd, i.e., $Z_2=1$, there always remains a helical mode pair which is protected by the Kramers theorem associated with the time-reversal symmetry. The phase transition between the trivial insulator ($Z_2=0$), and the topological insulator $(Z_2=1)$ occurs at the gap closing point, where the mass of the Dirac Fermion changes sign \cite{Bernevig2,Murakami2}.  

The existence of the QSH state has been predicted in semiconductors with an inverted electronic 
gap in HgTe/CdTe quantum wells \cite{Bernevig2}. 
The quantum well system experiences a quantum phase transition by changing the thickness.
%A generic quantum phase transition between QSH state and the insulator phases due to closing of the gap at the transition is discussed in Ref.\cite{Murakami2}.
Recent experiment has successfully demonstrated the existence of the helical edge mode for the quantum well of HgTe system by the measurement of the quantized charge conductance \cite{Konig}. 
Also recently, one can obtain the system very close to the quantum critical point by tuning the thickness of the quantum well. \cite{Mol}

Most of the previous works on QSH states have focused on properties of isolated QSH system, 
in particular, edge states of the system. However, in experiments to detect some characteristics 
of the QSH system, some probe, e.g. (metallic) electrode should be attached to the QSH. 
Therefore, it is an important issue to clarify transport property in N/QSH junctions. 
Moreover, the spin transport related to the QSH system is the most interesting issue, 
which has not been well explored and we will address in this paper. 
Especially, we focus on the spin transport properties \textit{normal} to the 
edge of the sample, while most of the previous works are interested in 
the charge transport \textit{along} the edge channel.   

In this paper, we study a reflection of the electronic wave at the N/QSH interface. 
It is found that an electron injected from the normal metal shows a spin dependent reflection at the interface leading to the spin rotation.
The spin rotation angle $\alpha$ shows rich structure centered around the 
quantum critical point $M=0$, and the normal incident angle $\theta=0$, 
i.e., it has a large value comparable to $\pi$ and even a winding by $4 \pi$ in the $(\theta-M)$-plane in the QSH region ($M<0$), which stems from the helical edge modes. 
This is in sharp contrast to the case of  normal metal/usual insulator interface. 
As an experimentally relevant system, we also investigate multiple reflections in QSH/N/QSH junction. It is clarified that the multiple reflection strongly enhances the spin rotation.

%The spin rotation angle is also much larger than that in the spin transistor proposed by Datta and Das \cite{Datta}, offering a quite efficient method for spin rotation.
%In QSH/N/QSH trilayer junction, multiple reflections at the interfaces even more enhances the spin rotation.

%%%%%%%%%%%%%%%%%%%%%%%%%%%%%%%%%%%%%%%%%%%%%%%%%%%%%%%%%%
% Formulation
%%%%%%%%%%%%%%%%%%%%%%%%%%%%%%%%%%%%%%%%%%%%%%%%%%%%%%%%
\begin{figure}[htb]
\begin{center}
\scalebox{0.8}{
\includegraphics[width=8cm,clip]{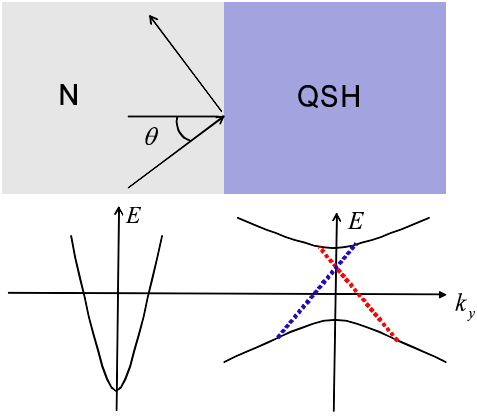}
}
\end{center}
\caption{(Color online) N/QSH junction and corresponding band structures (below). Dotted lines represent helical edge modes. 
}
\label{fig0}
\end{figure}

Let us commence with the the effective 4-band model proposed for
HgTe/CdTe quantum wells \cite{Konig2} 
\begin{eqnarray}
{\cal{H}}&=&\left(\begin{array}{cc} h(k)& 0\\
0&h^{*}(-k)\end{array}\right)\label{contH},
\end{eqnarray}
with $
h(k)=\epsilon (k) {\rm{I}}_{2\times 2}+d_a(k)\sigma^a,
\epsilon (k)=C-D(k_{x}^2+k_{y}^2), 
d_a (k)=\left(A k_x,-Ak_y,M(k)\right),
$ and
$M(k)= M-B(k_{x}^2+k_{y}^2)$ where we have used the basis order $({\vert
E_1 +\rangle,\vert H_1 +\rangle,\vert E_1 -\rangle,\vert H_1
-\rangle})$ ($"E"$ and $"H"$ represent the electron and hole bands, respectively), and,
$A,B,C,D,$ and $M$ are material parameters that depend on the quantum
well geometry.
%, and we have chosen the zero of energy to be the valence band edge of HgTe at ${\textbf{k}}=0$. 
${\cal{H}}$ is equivalent to 
two copies of the massive Dirac Hamiltonian but with a $k$-dependent mass $M(k)$.\cite{Konig2} 
This model is derived from the Kane model near the $\Gamma$
point in a quantum well of HgTe/CdTe junction, and its validity
is limited only near the $\Gamma$ point, namely for small values of $k_x$ and $k_y$. 
In this system, the transition of electronic band structure occurs from a normal to an 
inverted type when the thickness of the quantum well is varied through a critical thickness. 
This corresponds to the sign change of the mass $M$ of this system.\cite{Bernevig2,Zhou} 
%Note that in Ref. \cite{Zhou}, the above effective Hamiltonian is used to study finite size effects of helical edge states. In this paper, we use this Hamiltonian to investigate reflection problem at the N/QSH interface.

We consider the interface between normal metal and an insulator, the latter
of which is a usual insulator for $M>0$ and a QSH system for $M<0$, as shown in Fig. \ref{fig0}. For $M<0$, helical edge modes are expected to appear at the interface.\cite{Bernevig2,Zhou} 
The interface is parallel to $y$-axis and located at $x=0$, and the energy dispersion of each side is shown schematically in Fig. \ref{fig0}. The insulating side is described by the Hamiltonian given in Eq.(1). The Hamiltonian in the 
N side is given by setting $A =B = M = 0$ in that of QSH. 
The gap $|M|$ opens on the insulating side. $|C|$ is the Fermi energy measured from the bottom of the metallic band.

Now, let us focus on the spin up state, 
corresponding to the upper block of the Hamiltonian.  
Wave function in the N side for $E_1$ state injection is given by 
\begin{widetext}
\begin{equation}
\psi \left( {x \le 0} \right) = \left[ {\left( {\begin{array}{*{20}c}
   1  \\
   0  \\
\end{array}} \right)e^{ik_F \cos \theta x}  + r_E \left( {\begin{array}{*{20}c}
   1  \\
   0  \\
\end{array}} \right)e^{ - ik_F \cos \theta x}  + r'_E \left( {\begin{array}{*{20}c}
   0  \\
   1  \\
\end{array}} \right)e^{ - ik_F \cos \theta x} } \right]e^{ik_F \sin \theta y} 
\end{equation}
%\end{widetext}
with $k_F^2  = C/D$ and the incident angle $\theta$.

Wave function in the QSH side reads 
%\begin{widetext}
\begin{eqnarray}
\psi \left( {x \ge 0} \right) = \left[ {t\left( {\begin{array}{*{20}c}
   {Ak_ +  e^{i\theta _ +  } }  \\
   {d(k_ +  ) - M(k_ +  )}  \\
\end{array}} \right)e^{ik_ +  \cos \theta _ +  x}  + t'\left( {\begin{array}{*{20}c}
   {Ak_ -  e^{i\theta _ -  } }  \\
   {d(k_ -  ) - M(k_ -  )}  \\
\end{array}} \right)e^{ik_ -  \cos \theta _ -  x} } \right]e^{ik_F \sin \theta y} 
\end{eqnarray}
with $d(k) = \sqrt {A^2 k^2  + M(k)^2 }$ and 
\begin{eqnarray}
k_ \pm ^2  = \frac{1}{{2(B^2  - D^2 )}}\left[ { - (A^2  - 2MB + 2CD) \pm 
\sqrt {(A^2  - 2MB + 2CD)^2  - 4(B^2  - D^2 )(M^2  - C^2 )} } \right]
\end{eqnarray}
\end{widetext}
which is obtained by solving $E=C-D k^2+d(k)=0$ for $k$. Here, $\theta_\pm$ denote the 
incident angles for wavefunctions with wavenumber $k_\pm$. In the following, we set $C=0$ 
in the insulating side, when the Fermi energy is located at the middle of the gap. 
Notice that we assume insulating state so that $k_ \pm ^2<0$.
Due to the translational symmetry along the $y$-axis, we have 
$k_F \sin \theta  = k_ +  \sin \theta _ +   = k_ -  \sin \theta _ -  $. 

Boundary conditions read 
$ \psi \left( { + 0} \right) = \psi \left( { - 0} \right)$ and $v_x \psi 
\left( { + 0} \right) = v_x \psi \left( { - 0} \right)$ with $v_x  = \frac{{\partial H}}{{\partial k_x }}$. 
With these boundary conditions, we can obtain scattering coefficients.

Thus, the reflection coefficients for $E_1$ state injection are obtained as follows 
\begin{widetext}
\begin{eqnarray}
 \left( {\begin{array}{*{20}c}
   {r_E }  \\
   {r_E' }  \\
\end{array}} \right) = \frac{1}{{\Delta _E }}\left( {\begin{array}{*{20}c}
   {(Ak_ +  e^{i\theta _ +  }  - a_ +  )(d(k_ -  ) - M(k_ -  ) + b_ -  ) - (Ak_ -  e^{i\theta _ -  }  - a_ -  )(d(k_ +  ) - M(k_ +  ) + b_ +  )}  \\
   {2\left\{ { - b_ +  (d(k_ -  ) - M(k_ -  )) + b_ -  (d(k_ +  ) - M(k_ +  ))} \right\}}  \\
\end{array}} \right)
\end{eqnarray}
\end{widetext}
with $ \Delta _E  = (Ak_ +  e^{i\theta _ +  }  + a_ +  )(d(k_ -  ) - M(k_ -  ) + b_ -  ) - (Ak_ -  e^{i\theta _ -  }  + a_ -  )(d(k_ +  ) - M(k_ +  ) + b_ +  ),   a_\sigma   = \left( {1 + \frac{B}{D}} \right)A\lambda _\sigma  k_\sigma  e^{i\theta _\sigma  }  - A(d(k_\sigma  ) - M(k_\sigma  ))/(2Dk_F \cos \theta ), 
 b_\sigma   =  - A^2 k_\sigma  e^{i\theta _\sigma  } /(2Dk_F \cos \theta ) + \left( {1 - \frac{B}{D}} \right)\lambda _\sigma  (d(k_\sigma  ) - M(k_\sigma  ))$, 
 $ \lambda _\sigma   = k_\sigma  \cos \theta _\sigma /( k_F \cos \theta)$, and $\sigma  =  \pm $.

The corresponding reflection coefficients for spin down state can be obtained by the 
substitution $A \to -A$ and $\theta \to -\theta$. Spin rotation angle $\alpha$  is defined as $\alpha  = {\mathop{\rm Im}\nolimits} \log (r_ {E \uparrow } /r_{E \downarrow}  )$, 
which is the phase difference in the reflection coefficients between up and down spin electrons 
in the $E$ state. From the definition, we see that $\alpha$ is an odd function of $\theta$ 
since $r_{E \uparrow}(-\theta)=r_{E \downarrow}(\theta) $. 
Note that when the spin of the incident electron is within the $xy$-plane, this 
angle $\alpha$ gives the spin roration angle within this plane at the reflection in the same band. Therefore, we call it "spin rotation angle".  
%Note that the magnitude of $r_{E(H)}'$ is very small compared to $r_{E(H)}$, and hence we elucidate spin dependent reflection by $\alpha$ as defined above.

%%%%%%%%%%%%%%%%%%%%%%%%%%%%%% results
\begin{figure}[htb]
\begin{center}
\scalebox{0.8}{
\includegraphics[width=8.5cm,clip]{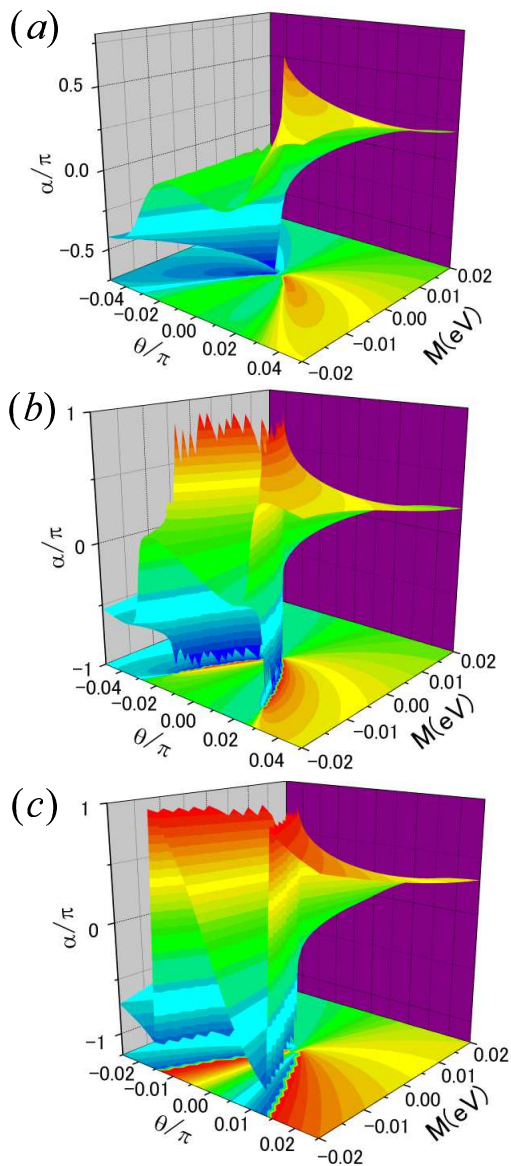}
}
\end{center}
\caption{(Color online) spin rotation angle $\alpha$ as a function of injection angle and $M$. 
(a) $C=-0.08$ eV, (b) $C=-0.1$ eV and (c) $C=-1$ eV. Note that $\alpha=\pi$ and $\alpha=-\pi$ are equivalent and the continuous increase of $\alpha$ in the clockwise direction is separated into several sheets in (b) and (c).
}
\label{fig1}
\end{figure}

%\begin{figure}[htb]
%\begin{center}
%\scalebox{0.8}{
%\includegraphics[width=8.0cm,clip]{fig2.eps}
%}
%\end{center}
%\caption{(Color online) magnitude of $\theta _C$, at which energy dispersion of the helical edge modes crosses the Fermi energy, as a function of $M$.
%}
%\label{fig2}
%\end{figure}

Now, we show the results for $\alpha$ in the plane of $(\theta,M)$ in Fig. \ref{fig1}
for $C=-0.08$ eV in (a), $C=-0.1$ eV in (b), and $C=-1$ eV in (c) with the other parameters fixed as $A=4$ $\rm eV \cdot \AA$, $B=-70$ $\rm eV \cdot \AA^2$ and  $D=-50$ $\rm eV\cdot \AA^2$.\cite{Konig2} 
In Figure \ref{fig1} (a), a sharp ridge in $M<0$ and $\theta>0$ region 
and its negative correspondence in $M<0$ and $\theta<0$ are seen.
This is in sharp contrast to the usual insulator case $M>0$ 
although $\alpha$ is still nonzero there.  Note that the
height of the ridge is as high as $\sim \pi/2$.
With increasing $|C|$, we find a qualitatively different structure. 
Near the origin in Fig. \ref{fig1} (b), $\alpha$ reaches $\pi$, changes its sign, and \textit{winds by $4 \pi$ around the origin}, while it does not in the region far away from the origin. 
One might wonder that the singularity occurs at the origin and there is 
also an endpoint of the "branch cut" separating 
the $4 \pi$ winding and no winding.
(Note that $\alpha=\pi$ and $\alpha=-\pi$ are equivalent 
and the continuous increase of $\alpha$ in the clockwise direction 
is separated into several sheets). 
In fact, we have checked that $\sin \alpha$, which is physically observable, depends on the direction from which the origin is approached but it does not show any singularity at the endpoint of the "branch cut".  As we further increase $|C|$,
the branch cuts extend toward the larger $|M|$ and $|\theta|$ region, and 
approaches gradually to the negative $M$-axis as shown in Fig. \ref{fig1} (c).
 Here, one may think that when magnetic field is applied in the $xy$ plane and opens a gap, it will change our results when the Fermi level is inside the gap. However, magnitude of the gap is typically smaller than that of $M (\sim$ 10meV), and the Fermi energy is shifted from the crossing of the helical edge dispersions. Hence, our results would be practically robust against applied magnetic field in realistic systems.
Note that all these interesting structures occur in the QSH region ($M<0$).
It should be also noticed that these structures are seen for small $\theta$'s, i.e., almost normal incidence, and the relevance to the helical edge channel 
is expected. 
%In fact, these features can be understood by the energy dispersions of the helical edge modes which cross the Fermi energy. 
The angle at which the dispersions of the helical edge modes hit the  Fermi energy is given by \cite{Zhou}
\begin{eqnarray}
\theta _C  = \pm \sin ^{ - 1} \left[ {\frac{{MD}}{{Ak_F \sqrt {B^2  - D^2 } }}} \right].
\end{eqnarray}
%Figure \ref{fig2} shows $|\theta _C|$ as a function of $M$ for several values of $C$, which corresponds to $k_F$. 
We find that a large magnitude of $\alpha$ in Fig. \ref{fig1} appears around this angle, which means that the helical edge modes resonantly enhance the spin rotation. This is plausible since in the helical edge modes, up and down spins propagate in an opposite direction and hence SU(2) symmetry in spin space is strongly broken there. 
This is also similar to the 
Andreev reflection in the superconducting analogue of QSH system 
where the helical edge channel produces a largely spin-polarized
supercurrent at the Andreev reflection. \cite{Tanaka}
%Therefore, it is concluded that this giant spin rotation angle is a unique feature to the N/QSH junction and to the quantum critical point corresponding to $M=0$. 

%When sign of $M$ changes, there occurs quantum phase transition between QSH state ($M<0$) and SHI ($M>0$). 
%Therefore, we find a \textit{qualitative} difference between these two phases, 
%In other words, the role of the helical edge modes on the scattering at the interface is rather 
%than qualitative.
%However, we find \textit{qualitative} difference between these phases when we shift the Fermi energy.

\begin{figure}[htb]
\begin{center}
\scalebox{0.8}{
\includegraphics[width=8.5cm,clip]{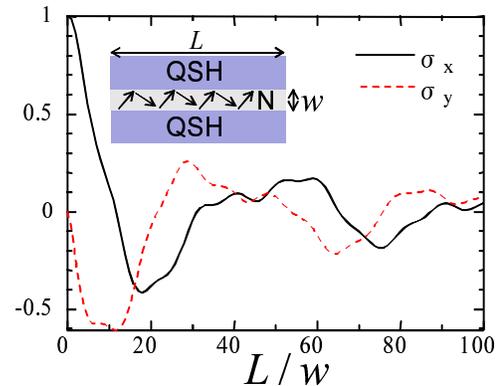}
}
\end{center}
\caption{(Color online) expectation values of $\sigma _x$ and $\sigma _y$  as a function of $L/w$ for $C=-1$ eV and $M=-0.01$ eV. Inset shows the model. 
}
\label{fig3}
\end{figure}

%%%%%%%%%%%%%%%%%%%%%%%%%%%%%%%%%%%%%%%%%%%%%%%%%%%%%%%%%%
% Results
%%%%%%%%%%%%%%%%%%%%%%%%%%%%%%%%%%%%%%%%%%%%%%%%%%%%%%%%
To realize the giant spin rotation experimentally, we propose QSH/N/QSH junction (see the inset in Fig. \ref{fig3}).  When N layer is sufficiently thin, the electrons in the N region are in close contact to the QSH systems and the transport property in this junction is determined by the reflection at the interfaces. 
%Thus, one can study the reflection property of QSH via transport property in this junction. 
We will also take statistical and incident angular average of the results in order to obtain more experimentally accessible quantity. 
%One may think that angular average prevents the spin rotation. However, we will show that  in this QSH/N/QSH junction, multiple reflections at the interfaces increase the phase difference between up and down spin states, and hence we can expect a remarkable spin rotation effect in this juction. 

%We assume the detector at the end of the wire which detects the component \eta  = (\eta _ \uparrow  ,\eta _ \downarrow  )^t. Then, we have
%\begin{eqnarray}
%\eta  \cdot \psi_N  = \eta _ \uparrow   \cdot R_ \uparrow ^N \psi _ \uparrow   + \eta _ \downarrow   \cdot R_ \downarrow ^N \psi _ \downarrow.
%\end{eqnarray}

%The intensity is defined as
%\begin{eqnarray}
%P = \left\langle {\left| {\eta  \cdot \psi_N } \right|^2 } \right\rangle 
%\end{eqnarray}
%where $\left\langle ... \right\rangle$ denotes angular averaging.

To investigate spin rotation effect in this junction, we also have to calculate  reflection coefficients with $H$ state injection and $E$ or $H$ state reflection in a similar way. 
We consider the initial wavefunctions with density matrix as
\begin{eqnarray}
\psi _i  = \left( {\begin{array}{*{20}c}
   {a_1 }  \\
   {a_2 }  \\
   {a_3 }  \\
   {a_4 }  \\
\end{array}} \right),\;\overline {a_i^* a_j }  = \left\{ \begin{array}{l}
 1/4\quad {\rm{for}}\left\{ {i,j} \right\}\, = \left\{ {1,3} \right\},\left\{ {2,4} \right\} \\ 
 0\quad {\rm{otherwise}} \\ 
 \end{array} \right. \label{ini}
\end{eqnarray}
where $\overline {...}$ denotes statistical average. 
The density matrix is chosen so that the initial wavefunctions have a correlation in the same bands.  As seen in Eq. (\ref{ini}), here we consider the initial wavefunctions with spin polarization along $x$-axis. 
Using these expressions, we can obtain expectation values of $\sigma _x$ and $\sigma _y$, Pauli matrices in spin space.

%%%%%%%%%%%%%%%%%%%%%%%%%%%%%%%%%%%%%%%%%%%%%%%%%%%%%%%%%%
% Results
%%%%%%%%%%%%%%%%%%%%%%%%%%%%%%%%%%%%%%%%%%%%%%%%%%%%%%%%

Figure \ref{fig3} displays expectation values of $\sigma _x$ and $\sigma _y$ as a function of $L/w$ for $C=-1$ eV and $M=-0.01$ eV. Here, $L$ and $w$ are the length and width of the N, respectively. 
An oscillatory dependence on $L/w$ is seen because the phase difference between up and down spin states increases with $L/w$, namely the number of reflections. When $\sigma_x$ has an extremum, $\sigma _y$ almost vanishes and vice versa, which indicates a rotation of the spin in the $xy$ plane in spin space upon propagation in the N of QSH/N/QSH junction. This means that the predicted giant spin rotation persists and is observable in realistic systems. 
Around $L/w=20$, the initial spin is rotated by $\pi$. With $w=$5nm, this giant spin rotation can be realized with the very short length scale of $L=$100nm, which should be compared with that in the previous work\cite{Datta} where the propagation of electron over 1$\mu$m is required to rotate electron spin by $\pi$. This may be a great advantage for application to nanotechnology.

In summary, we studied a reflection problem at the N/QSH interface and showed that an electron 
injected from the normal metal in N/QSH junctions shows a spin dependent reflection at the interface and hence there appears a phase difference between 
up and down spin states in the reflection coefficients. 
%This effect is quite remarkable compared to that in N/SHI junction: 
The spin rotation angle can be as large as $\sim \pi$ in the QSH junction, because the helical edge modes resonantly enhance the spin rotation.  
In the QSH/N/QSH junction, multiple reflections at the interfaces increase the phase difference. This also results in a remarkable spin rotation effect in this juction even when the results are averaged over incident angles.
The proposed Fabry-P\'erot like heterostructure is experimentally accessible and could be used to unveil another aspect of the QSH state: \textit{spin rotation effect}. 

%Moreover, in contrast to the spin transistor,\cite{Datta} the system we considered has a much stronger spin rotation effect. 
%In Ref. \cite{Datta}, the junction length $\sim$ 1$\mu$m is required to rotate electron spin by $\pi$ but this can be achieved with much shorter junctions in our theory since $\pi$ phase shift can occur by only one reflection, which may be a great advantage for application to nanotechnology.

%Our prediction gives an important example for a possible realization of spin transport without dissipation.\cite{murakami2004a,Tanaka}

This work is supported by Grant-in-Aid for Scientific Research (Grant No.
17071007, 17071005) %on Priority
%Area "Novel Quantum Phenomena Specific to Anisotropic Superconductivity"
%(Grant No. 17071007) and B (Grant No. 17340106)
from the Ministry of Education, Culture, Sports, Science and Technology of
Japan and NTT basic research laboratories. 
T.Y. acknowledges support by JSPS.

%---------------------

\end{document}